\documentclass{article}

\usepackage{spconf,amsmath,graphicx}
\usepackage{subcaption}
\usepackage{amsfonts}
\usepackage{mathtools}
\usepackage{bm}
\usepackage{hyperref}
\usepackage{booktabs}
\usepackage{tabularx}

\usepackage{xcolor}

\usepackage{multirow}

\pdfoutput=1

\title{End-to-end Non-Negative Autoencoders for Sound Source Separation}

\name{Shrikant Venkataramani$^{\sharp}$ \qquad 
      Efthymios Tzinis$^{\flat}$ \qquad 
      Paris Smaragdis$^{\flat, \natural,}$\sthanks{Supported by NSF grant \#1453104}}

\address{$^{\sharp}$University of Illinois at Urbana-Champaign, Department of Electrical and Computer Engineering \\
         $^{\flat}$University of Illinois at Urbana-Champaign, Department of Computer Science \\
         $^{\natural}$Adobe Research}

\begin{document}
\ninept
\maketitle
\begin{abstract}
Discriminative models for source separation have recently been shown to produce impressive results. However, when operating on sources outside of the training set, these models can not perform as well and are cumbersome to update. Classical methods like Non-negative Matrix Factorization (NMF) provide modular approaches to source separation that can be easily updated to adapt to new mixture scenarios. In this paper, we generalize NMF to develop end-to-end non-negative auto-encoders and demonstrate how they can be used for source separation. Our experiments indicate that these models deliver comparable separation performance to discriminative approaches, while retaining the modularity of NMF and the modeling flexibility of neural networks.
\end{abstract}

\section{Introduction}
\label{sec:intro}
Given a mixture of multiple concurrent sources, the aim of single-channel source separation is to extract the individual sources from the mixture. Under a supervised training setup, we assume that we have clean training examples for each source in the mixture. Using these training examples, we first construct suitable models for the sources. Then, we use these trained models to extract the in sources from previously unseen mixtures. 

With the current rapid strides in neural networks (NNs) and deep learning, several sophisticated architectures have been proposed and successfully used for single-channel source separation~\cite{hershey2016deepclustering, venkataramani2018end, huang2015jointmasksoptimizationDRNNs, jansson2017singing}. More recently, we have started to operate directly on the waveforms with several end-to-end approaches available~\cite{venkataramani2018end, luo2018tasnet, le2019phasebook}, and use better cost-functions motivated by the Source-to-Distortion ratio (SDR)~\cite{vincent2006performance, leglaive2019speech, venkataramani2018performance, mimilakis2018examining, kolbaek2019loss, venkataramani2018end}. Using deep-clustering~\cite{hershey2016deepclustering} and permutation-invariant training~\cite{Yu2017PIT}, we can train the networks to perform speaker-independent source separation. These advancements have resulted in significant improvements in separation performance. 

However, these neural network models are typically trained under a discriminative setup. The mixture is given as an input to the network and the network is trained to produce outputs that resemble the constituent clean sources. Thus, for these models to work well in the real world, we need huge networks and copious amounts of training examples that account for different signal-to-noise ratio combinations and various categories of sounds. In such cases, updating trained networks to operate on mixtures containing a new sound requires significant data-augmentation and re-training on the augmented dataset. This is particularly infeasible when operating on a wide variety of sources~\cite{kavalerov2019universal}, especially with limited computational resources.

In contrast, classical approaches like Non-negative Matrix Factorization (NMF)~\cite{smaragdis2003non, virtanen2007monaural, fevotte2009nonnegative, fevotte2011algorithms, ozerov2009multichannel} and its recent NN alternatives~~\cite{smaragdis_alternatives_nonnegative_models, venkataramani_alternatives_conv_models} use generative models for source separation. We use clean training examples to build models for the sources and use these models to estimate the contribution of the sources in unseen mixtures. A significant advantage of such generative models is that the trained models can be used to extract the corresponding source from any mixture irrespective of the interfering sources in the mixture. Thus, tweaking the models to work for a new type of source just requires training a new model for the source and appending it to the existing dictionary of models. However, these approaches continue to lag behind discriminative models due to a lack of updates to their capabilities. We aim to address this issue here.

In this paper, we improve upon Non-Negative Autoencoders (NAEs) and generative models for source separation. We generalize NAEs to develop end-to-end versions and show how they can be used for single-channel source separation. These networks simultaneously operate on custom-designed time-frequency representations that are optimal for each source. With these improvements, we demonstrate that end-to-end NAE networks are comparable to discriminative approaches in terms of source separation performance. At the same time, we retain the advantages of generative modeling. The models are reusable and can be used in a variety of test scenarios without any additional data-augmentation or training requirements.


\section{End-to-end Non-Negative Autoencoders}
\label{sec:end_to_end_modular_networks}
We first begin with a brief description of NAEs. We extend the capabilities of NAEs to operate directly on the audio waveforms to construct end-to-end NAE networks.

\subsection{Non-negative Autoencoders}
\label{ssec:non_negative_autoencoders}
NMF approximates a non-negative matrix $\mathbf{S} \in \mathbb{R}_{\geq 0}^{M \times N}$ as a product of two low-rank non-negative matrices, $\mathbf{W} \in \mathbb{R}_{\geq 0}^{M \times K}$ and $\mathbf{H} \in \mathbb{R}_{\geq 0}^{K \times N}$. Here, $\mathbb{R}_{\geq 0}^{M \times N}$ denotes the set of matrices size $M \times N$ with real, non-negative elements. In the case of audio signals, we apply NMF on the magnitude spectrograms. The columns of $\mathbf{W}$ (NMF bases) learn spectral bases and the rows of $\mathbf{H}$ (NMF activations) indicate the weights of the bases in each frame of the spectrogram.

As described in~\cite{smaragdis_alternatives_nonnegative_models}, we can generalize these NMF models by interpreting them as a neural network. In the case of NMF, we can replace it by a two-layer NAE given by, 

\begin{align}
    \text{$1^{\text{st}}$ layer:~ (Encoder)~}\mathbf{H} &= g(\mathbf{W^{\ddagger}} \cdot \mathbf{S}) \nonumber \\
    \text{$2^{\text{nd}}$ layer:~ (Decoder)~}\mathbf{\widehat{S}} &= g(\mathbf{W} \cdot \mathbf{H})
    \label{eq:nmfae}
\end{align}

In this equation, $\mathbf{S}$ represents the input spectrogram, the decoder weights $\mathbf{W}$ gives the equivalent of NMF bases and the encoder output $\mathbf{H}$ gives the equivalent of the NMF activations. The weights of the encoder $\mathbf{W^{\ddagger}}$ represent a form of a pseudo-inverse matrix that can be applied to the input spectrogram to get the activations. The non-linearity $g(.):\mathbf{R}\rightarrow \mathbf{R}_{\geq 0}$ operates element-wise and maps a real number to the space of positive-real numbers. This allows the network to learn a non-negative $\mathbf{H}$ and a non-negative reconstruction $\mathbf{\widehat{S}}$ of the input $\mathbf{S}$. Unlike NMF, the decoder weights need not be strictly non-negative. But, under suitable sparsity constraints on the activations $\mathbf{H}$, they can be shown to be non-negative like NMF bases~\cite{smaragdis_alternatives_nonnegative_models, venkataramani_alternatives_conv_models}. 

Eq.~\ref{eq:nmfae} defines the encoder and decoder of our NAE to be a single dense-layer. However, we are not necessarily restricted by this formulation. We can now take advantage of the modeling flexibility of neural networks and develop complex encoder and decoder architectures that adhere to the above format. In particular,  multi-layer~\cite{smaragdis_alternatives_nonnegative_models} and convolutional extensions~\cite{venkataramani_alternatives_conv_models} have shown significant performance improvement compared to a single dense-layer encoder and decoder given in Eq.~\ref{eq:nmfae}. Probabilistic equivalents of these models using variational NAEs have also been proposed~\cite{squires2019variational, karamatli2019weak}. As before, the weights of the decoder act as a representative model for the source. The output of the encoder indicates the corresponding activations of the model to explain the input to the autoencoder.

\subsection{End-to-end Processing}
\label{ssec:end_to_end_processing}

In addition to using complex architectures, recent neural network approaches for source separation operate on the mixture waveforms and estimate the waveforms of the constituent sources directly. Adopting such end-to-end approaches has yielded a significant boost in separation performance using neural networks~\cite{venkataramani2018end, luo2018tasnet}. Even in the case of NMF, learning a front-end transform has led to significant improvements~\cite{fagot2018nonnegative, wendt2018jacobi}.

\begin{figure}[!h]
    \centering
      \begin{subfigure}{0.34\linewidth}
      \flushleft
          \includegraphics[width=\linewidth]{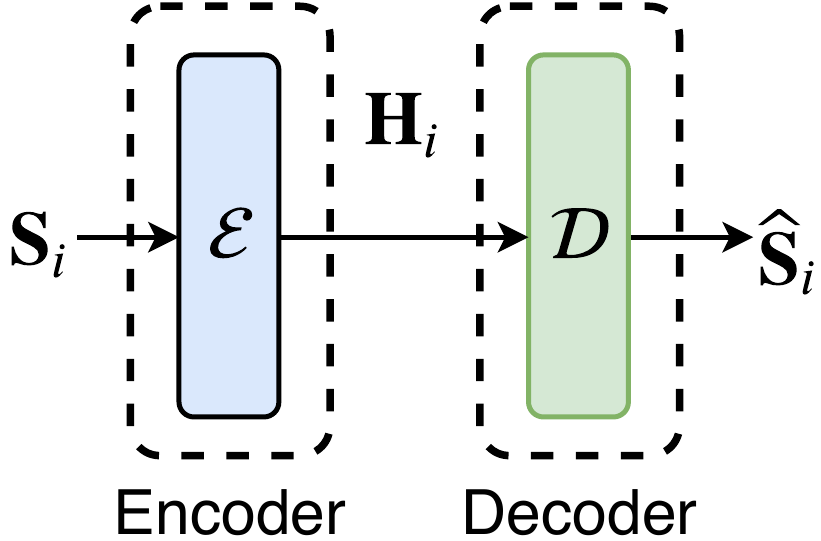}
          \caption{}
          \label{fig:training_simple_NAE}
      \end{subfigure} \hspace{6mm}%
      \begin{subfigure}{0.56\linewidth}
      \centering
          \includegraphics[width=\linewidth]{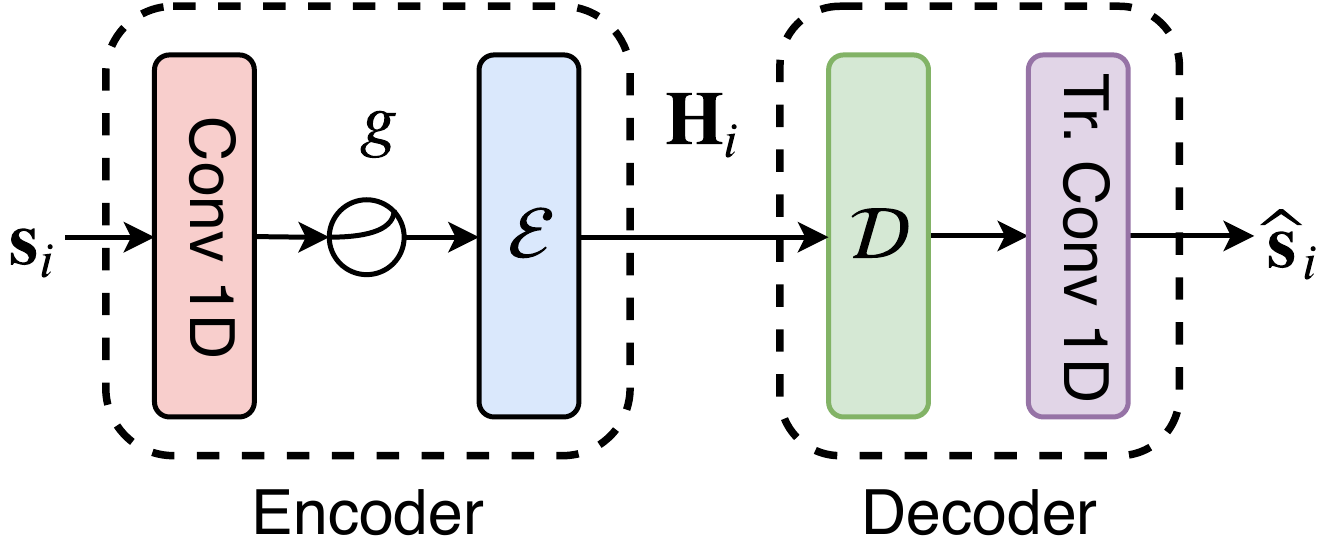}
          \caption{}
          \label{fig:training}
      \end{subfigure}
    \caption{Block diagrams for: a) a non-negative autoencoder (NAE) b) an end-to-end NAE. Here, $\mathcal{E}$ and $\mathcal{D}$ represent the encoder and decoder of the NAE respectively. We append a 1D-convolutional layer as a front-end and back-end to enable the network operate to on waveforms directly. Thus, the end-to-end NAE encoder consists of the front-end layer and the NAE encoder $\mathcal{E}$. Similarly, end-to-end NAE decoder is made up of the NAE decoder $\mathcal{D}$ and the back-end layer. In the training step, we build and train an end-to-end NAE for every source we hope to encounter. The trained model is then used in the inference step for separating the sources.}
    \label{fig:all_training}
\end{figure}

To introduce end-to-end processing capabilities into our NAE, we replace the front-end transform step by a 1D-convolutional layer. To get a non-negative representation like the magnitude spectrogram, we use a softplus non-linearity for the layer. To transform back into the waveform from the latent representation, we use a transposed 1D-convolutional layer. These modifications allow the network to accept a waveform and learn a trainable latent representation that is optimal for representing a particular source. As we will show in Section~\ref{sec:supervised_source_separation}, this also enables operating simultaneously on multiple customized latent representations corresponding to the different sources in the mixture, to extract the sources. Fig.~\ref{fig:training} shows the block diagram of our end-to-end NAE.
\section{Supervised Source Separation}
\label{sec:supervised_source_separation}
Having developed the end-to-end NAE architecture, we now show how we can use it for end-to-end source separation. Like NMF, source separation using end-to-end NAEs is a two-step process,\\
\underline{\textit{Step 1}}: Learn suitable end-to-end NAE models for all the sources we expect to encounter in the mixture. We refer to this as the ``training'' step. \\
\underline{\textit{Step 2}}: Given an unseen mixture, fit the trained models to explain the contributions of the individual sources in the mixture. We refer to this as the ``inference'' step.\\

\begin{figure}[ht]
    \centering
  \begin{subfigure}{0.36\linewidth}
  \flushleft
      \includegraphics[width=\linewidth]{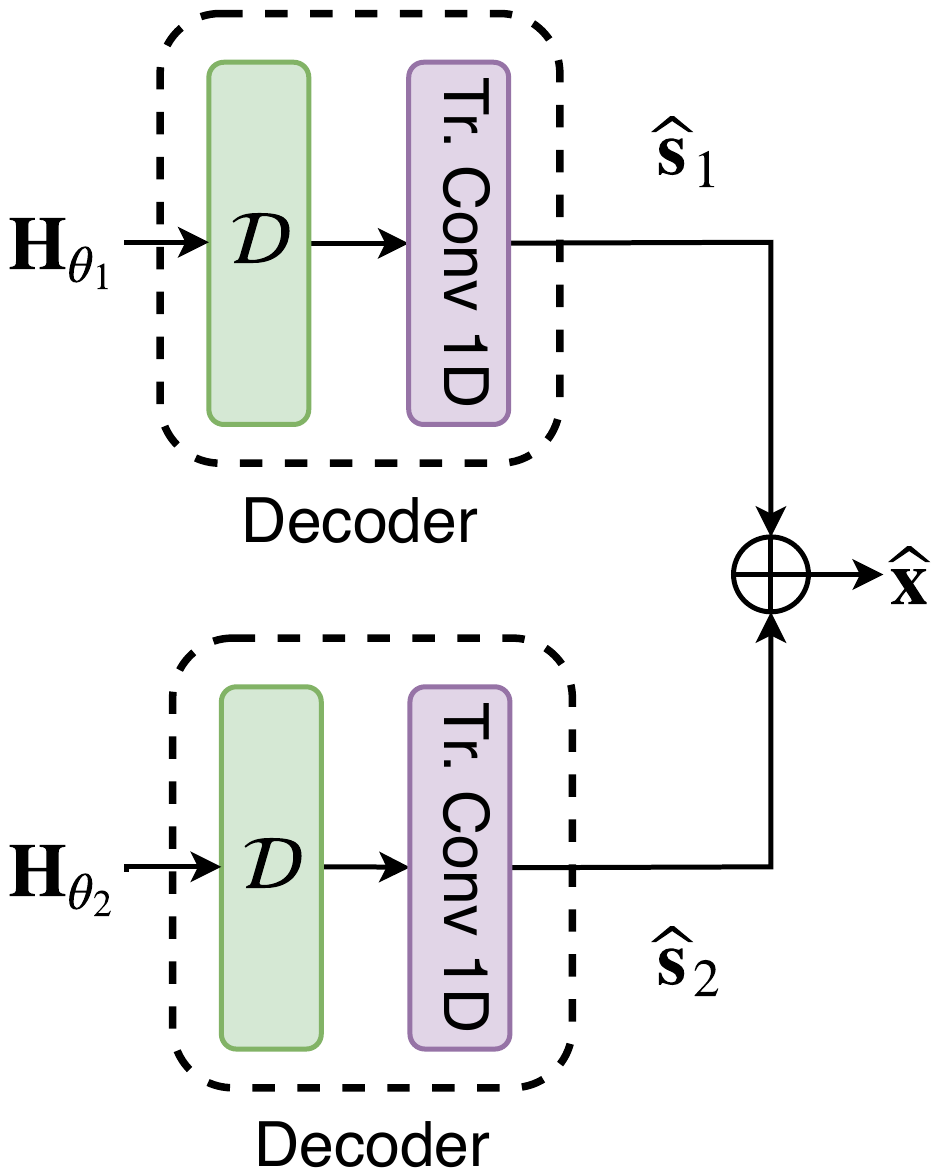}
      \caption{}
      \label{fig:inference_latent}
  \end{subfigure} 
  \begin{subfigure}{0.63\linewidth}
  \centering
      \includegraphics[width=0.98\linewidth]{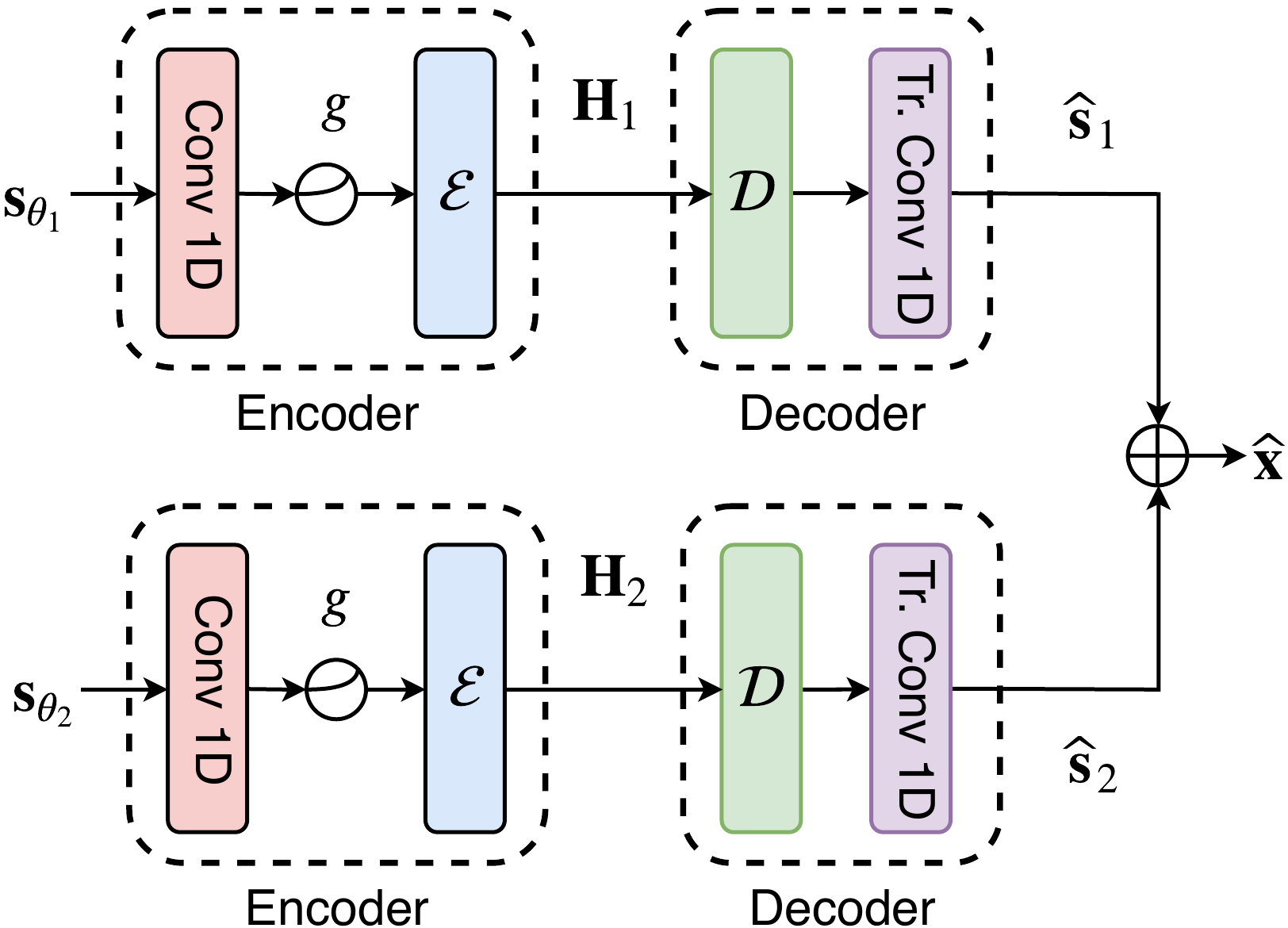}
      \caption{}
      \label{fig:inference_time}
  \end{subfigure}
    \caption{Block diagram of source separation using end-to-end NAEs during the inference step: estimating the model activations $\mathbf{H_{\theta_{i}}}$ of the sources (left) and estimating the time-domain waveforms of the sources $\mathbf{s_{\theta_{i}}}$ (right). The subscript $\mathbf{\theta}$ indicates that the corresponding variable is a parameter of the inference network and we train to estimate its values for the given mixture.}
    \label{fig:inference}
\end{figure} 

To extract the sources, we use the pre-trained end-to-end NAEs from the training step to construct an inference network. In this paper, we consider two distinct inference frameworks. As shown in~\cite{smaragdis_alternatives_nonnegative_models}, the decoders of the pre-trained NAEs sufficiently act as representative models. Extending the same idea, the decoders of our pre-trained end-to-end NAEs can be used to separate sources in the inference step. It is important to emphasize here that when we refer to the ``decoder'' in case of end-to-end NAEs, it contains the NAE decoder~$\mathcal{D}$ followed by the transposed-1D convolution layer. The block diagram of the inference network using the pre-trained end-to-end decoders is shown in Fig.~\ref{fig:inference_latent}. At the outputs of the decoders, we can directly access the waveforms of the individual sources. Consequently, the goal of the inference step is to extract the source waveforms $\mathbf{s_{i}}$ for $i \in {1,2, \dots K}$, given the mixture waveform $\mathbf{x_{m}}$ and the pre-trained decoders. To get these decoder outputs, we need to estimate the activations $\mathbf{H_{\theta_{i}}}$ for all the sources in the mixture. In other words, during the inference step, we optimize towards finding the right inputs to the inference network. The subscript $\mathbf{\theta}$ serves to denote the trainable parameters of our inference network.

The above inference procedure ignores the encoder in constructing the inference network. Incorporating the pre-trained encoder could potentially improve separation performance. Thus, we consider an alternative approach where the inference network uses the whole pre-trained end-to-end NAE. As before, we optimize to find the right inputs~$\mathbf{s_{\theta_{i}}}$ to the inference network such that the outputs add up to explain the mixture. In this version of inference, we are optimizing on the space of waveforms as opposed to the space of activations. Fig.~\ref{fig:inference_time} describes this inference approach as a block diagram. 

To develop an intuition on the complexity of the inference optimization, we can compare the number of trainable parameters for the two inference approaches for a $1$-second test example. For a sampling rate of $16$ kHz, a $64$ dimensional activation matrix for each source and a stride of $32$ samples for the front-end 1D-convolutional layer, the number of trainable parameters can be given by $16,000 \cdot \frac{64}{32} = 32,000$ parameters per source. The second inference approach optimizes to estimate the waveforms as the parameters and trains for $16,000$ parameters per source. Thus, the inference optimization happens on a significantly smaller parameter space compared to training standard neural networks. In addition, applying the inference step for longer test examples can possibly be done in batches, reducing the inference time further.

\subsection{Cost-function}
\label{ssec:cost_function}
In the case of discriminative end-to-end models, using cost-functions motivated by SDR have led to several promising results~\cite{venkataramani2018performance, luo2019convTasNet}. We can use these waveform based cost-functions to train end-to-end NAEs, both during training and inference. For a reference waveform $\mathbf{y}$ and a network output $\mathbf{x}$, we maximize a simplified version of SDR given by $\frac{|\left<\mathbf{x},~\mathbf{y}\right>|^2}{\left<\mathbf{x},~\mathbf{x}\right>}$. Here, $\left<\mathbf{x},~\mathbf{y}\right>$ represents the inner-product operation. Intuitively, we ask to maximize the sample correlation between $\mathbf{x}$ and $\mathbf{y}$ and while minimizing the energy of the solution. 

\subsection{Advantages of end-to-end NAEs}
\label{ssec:advantages}
We now consider the advantages that end-to-end NAEs have to offer when used for source separation. As stated previously, we can exploit the modeling flexibility afforded by neural networks to construct complex architectures that operate on the waveforms directly. In our experiments, we show that end-to-end NAE models are comparable to discriminative models in terms of separation performance. Although we require an optimization process during inference, we gain a significant advantage. End-to-end NAEs are developed as generalizations of NMF and we continue to retain its modular nature. Consequently, once we learn a model for a source, we can use the model on any mixture that contains the source and extract it, irrespective of the interferences in the mixture. Also, we can directly use the pre-trained models without the need for any data-augmentation, on a variety of test examples with varying characteristics. We also evaluate this capability of end-to-end NAEs in our experiments. In fact, the availability of an inference step where, we try to optimally fit the pre-trained models on an unseen mixture, allows us to use the same models for different test mixtures. In other words, the modularity of end-to-end NAEs is a consequence of having a trainable inference step. Finally, extending the model to operate on new types of sounds becomes extremely easy. All we need to do is to train an end-to-end NAE for the new source. We can then append the pre-trained model to the inference network to separate that source from given mixtures.


\section{Experiments}
\label{sec:experiments}
Having described the training and inference steps, we now present some experiments and their results to evaluate the performance of end-to-end NAEs for supervised single-channel source separation. Primarily, we focus on two experiments. The first experiment is aimed at comparing end-to-end NAEs to end-to-end discriminative source separation models, in terms of their separation performance. The second experiment is directed towards evaluating their modular nature. We begin with a description of our experimental setup.

\subsection{Dataset}
\label{ssec:dataset}
For our experiments, we use the Device and Produced Speech~(DAPS) dataset~\cite{mysore2014can}. We use the clean speech examples from the dataset only, for our experiments. Of the $10$ male and $10$ female speakers, we use $8$ male and $8$ female speakers to construct the training set. We use the first $3$ scripts to construct the training examples out of the $5$ scripts available per speaker.  This gives about $2$ hours of training data for each class. To evaluate separation performance, we generate two test sets as follows. \textit{Test-set-1} is generated from the unused recordings of the speakers that are a part of the training set. The test examples in \textit{Test-set-2} come from the speakers not included in the training set. We down-sample the recordings to $16$ kHz and randomly draw $2$-sec snippets for training and testing.

\subsection{Network}
\label{ssec:network_config}
For the network configurations, we use a front-end 1D convolutional layer consisting of $256$ filters of width $64$ samples and a stride of $32$ samples. The NAE encoder ($\mathcal{E}$) is formed by a cascade of two 1D-convolutional layers. The two layers have $128$ and $64$ filters respectively, a filter width of $5$ taps and a stride value of $1$ respectively. Thus, the activation matrix for each source has a dimensionality of $64$. The decoder architecture is constructed to invert the activation matrix. Thus, the NAE decoder ($\mathcal{D}$) is constructed using two 1D transposed-convolutional layers of $128$ and $256$ filters, a filter width of $5$ taps and a stride value of $1$. Each convolutional layer is followed by a softplus non-linearity and a batch-norm operation. To transform the output of the NAE decoder $\mathcal{D}$ back into the waveform domain, we use a 1D transposed-convolutional layer having the same parameters as the front-end. In our experiments, we compare the performance of NAEs to discriminatively trained source separation models. For the discriminative model, we use the same architecture as an end-to-end NAE.

\begin{figure*}[ht]
    \centering
  \begin{subfigure}{0.45\linewidth}
  \flushleft
      \includegraphics[width=\linewidth]{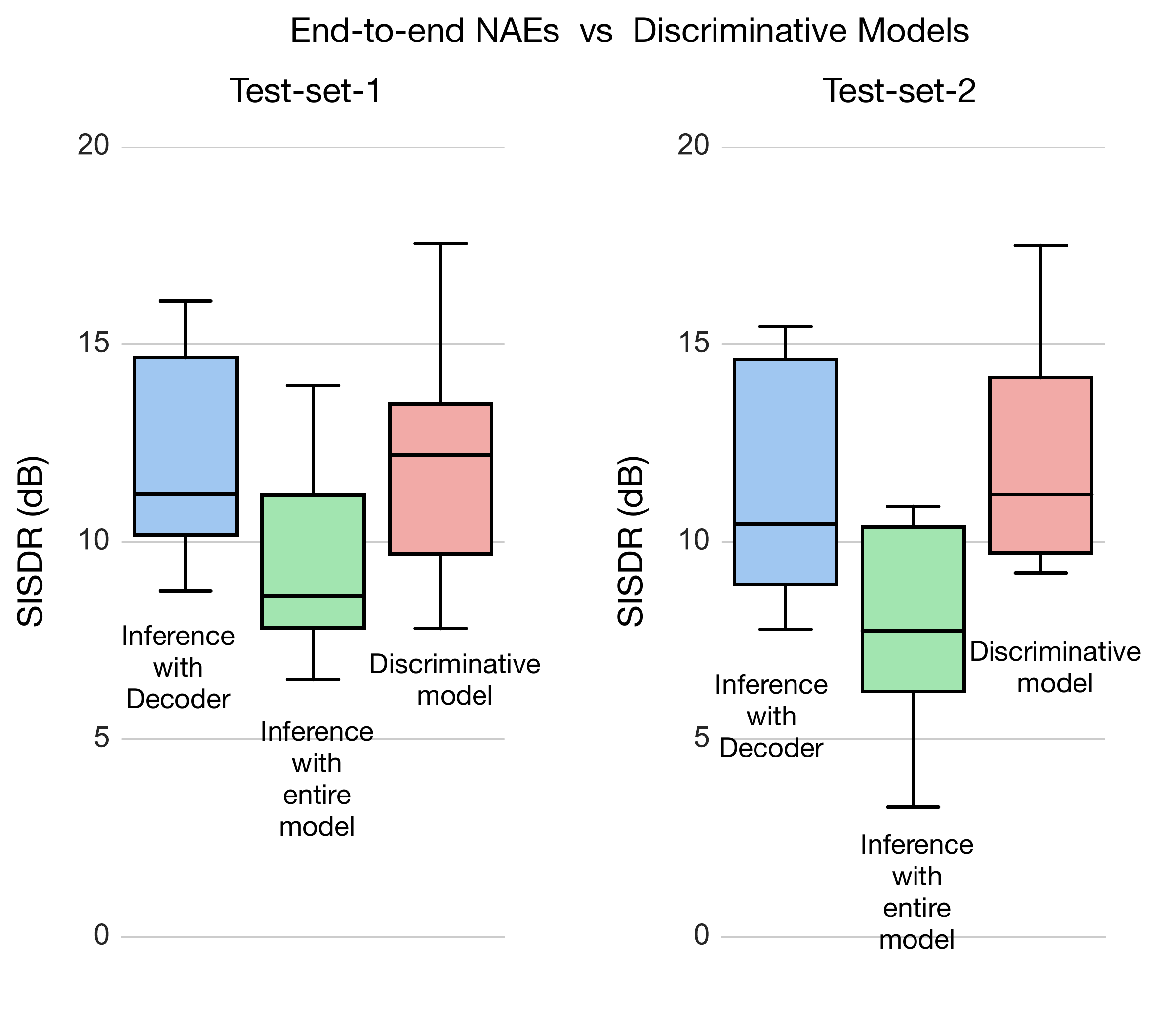}
      \caption{}
      \label{fig:exp1}
  \end{subfigure} \hspace{12mm}%
  \begin{subfigure}{0.45\linewidth}
  \centering
      \includegraphics[width=\linewidth]{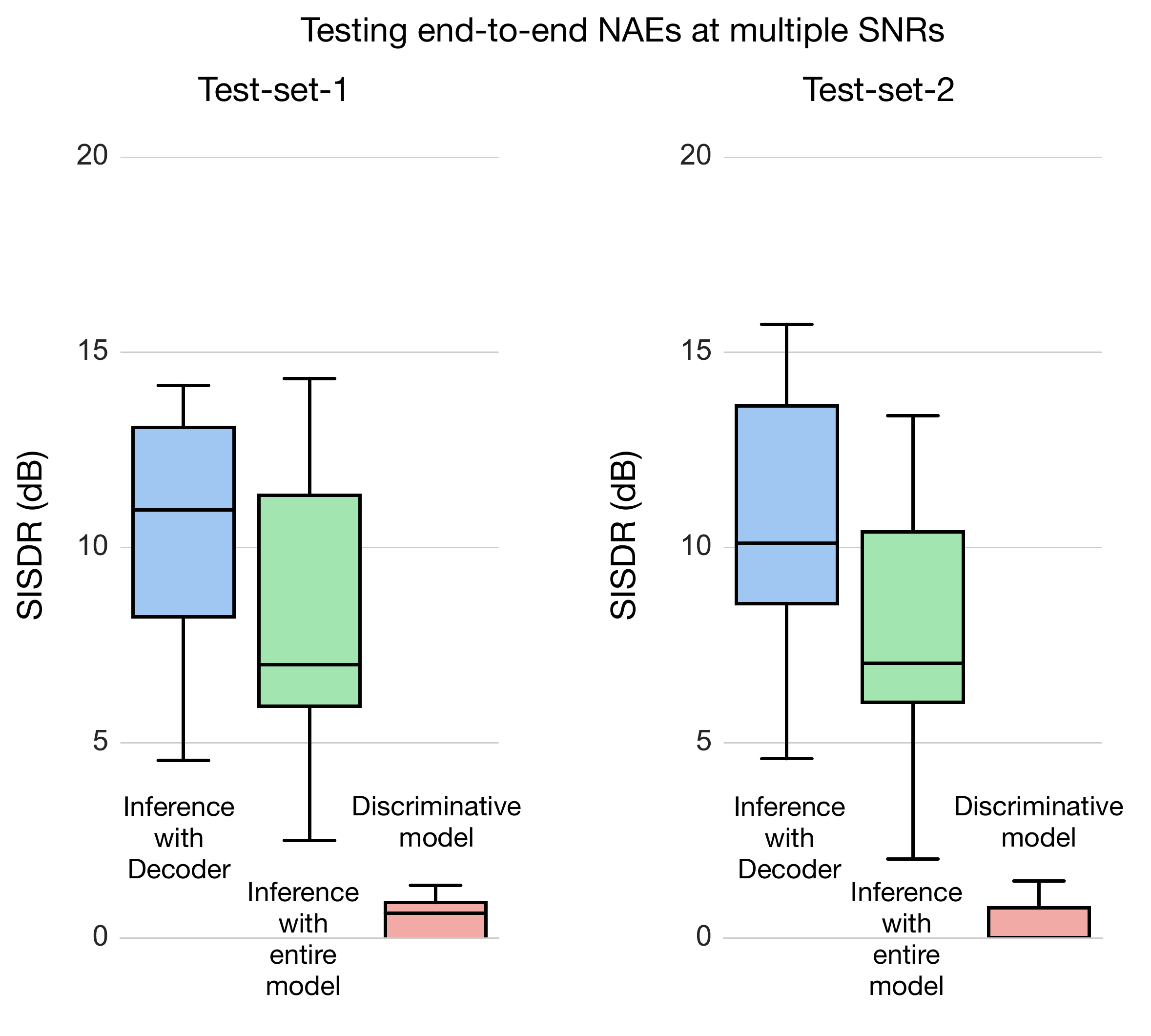}
      \caption{}
      \label{fig:exp2}
  \end{subfigure}
    \caption{(a) SISDR values for $0$ dB test mixtures. We see that the separation performance of end-to-end NAEs is comparable with discriminatively trained models. (b) SISDR values for test mixtures with SNR varying between $-3$ dB and $3$ dB. We see that the inference step allows us to fit the models and separate sources even for conditions unseen during the training step. The discriminative model cannot deal with this mismatch between the training and test sets. Test-set-1 consists of test examples whose speakers were a part of the training set. Test-set-2 is drawn from speakers not included in training. The distribution of SISDR values is shown in the form of a box-plot. The solid line in the center of the box shows the median value and the box-boundaries show the inter-quartile range ($25^{th}$ and $75^{th}$ percentile points). }
    \label{fig:exp}
\end{figure*} 

\subsection{Results and Discussion}

\subsubsection{End-to-end NAEs v/s Discriminative models}
The first experiment is designed to compare the separation performance of end-to-end NAEs with discriminative source separation models. We train two end-to-end NAEs, one for the set of male speakers and the other for the set of female speakers. Using these pre-trained models, we apply the inference step on the test mixtures to get the separation results. The discriminative model used for comparison is trained as a denoising auto-encoder that separates the female speaker from a $0$ dB mixture consisting of a male and a female speaker. We generate these $0$ dB training examples by drawing random snippets from the training sets and mixing them. For evaluation, we generate $30$ test mixtures mixed at $0$ dB for each test set. To reiterate, Test-set-1 is formed from speakers that are included in the training set. Test-set-2 is generated from speakers that are not included in the training set. The results are shown in Fig.~\ref{fig:exp1} as a box-plot of scale-invariant SDR (SISDR)~\cite{le2019sdr, vincent2006performance} values.

As shown in Fig.~\ref{fig:exp1}, we compare the separation performance of three models: 1. End-to-end NAE with inference using decoder only (left) 2. End-to-end NAE with inference using entire model (middle) and 3. End-to-end discriminative source separation (right) over both the test-sets. We see that the separation performance of end-to-end NAEs with decoder based inference is comparable with end-to-end discriminative models over both the test sets. Using the entire model for inference results in a significant drop in separation performance. This can be attributed to the fact that inference on waveforms does not produce silences as effectively as inference on the activation matrices. Comparing with the performance on Test-set-2, we observe a dip in median SISDR on end-to-end NAE models. Also, the variance in SISDR values increases slightly. But, the separation performance looks about the same overall.

\subsubsection{Testing end-to-end NAEs at multiple SNRs}

The second experiment aims to evaluate the modular nature of our end-to-end NAEs. We evaluate this by comparing the performance on mixtures with varying signal-to-noise ratio (SNR) levels. In the case of discriminative models, we cannot expect any reasonable separation results when there is a mismatch between the training and test mixtures encountered by the models. But, the model-fitting performed during the inference step for end-to-end NAEs allow the use of the pre-trained models even in mismatched cases. We generate $30$ test mixtures with SNR levels varying from $-3$ dB to $3$ dB, taking the male speech as the reference. As before, we construct two test sets of $30$ mixtures each: test-set-$1$ from speakers included in the training set and test-set-$2$ from speakers that are not a part of the training examples. We compare across three models: 1. End-to-end NAE with decoder inference (left) 2. End-to-end NAE with fill model inference (middle) and 3. End-to-end discriminative source separation (right). We use the same models trained for the previous experiment in all the cases. The results are shown in Fig.~\ref{fig:exp2}

We see that the end-to-end NAEs achieve a good separation even at these varying SNR levels, something not possible in the case of discriminative separation models as shown by their extremely poor performance. However, we also observe an increase in the range and variance of SISDR values for this experiment. As before, we see a drop in the median SISDR values and an increase in the variance, on Test-set-2 compared to Test-set-1. Despite this, the separation performance falls in the same range overall, indicating the efficacy of the trained model.

\section{Conclusion}
\label{sec:conclusion}
In this paper, we developed and investigated the use of end-to-end non-negative autoencoders for supervised source separation. These networks were developed as a generalization of NMF and can learn suitable models for the sources directly from their waveforms. Our experiments showed that these networks are comparable to discriminative source separation models in terms of their separation performance. Although these models exhibit a more computationally intensive inference step, they allow for additional extensions that discriminative models cannot easily facilitate. In addition to having an extensible architecture, the modular nature of end-to-end NAEs allowed us to separate sources from mixtures having a wide range of signal-to-noise ratios using the same pre-trained generative models for the sources.


\bibliographystyle{IEEEbib}
\bibliography{mybib}
\end{document}